\documentclass[prl,twocolumn,showpacs,preprintnumbers,amsmath,amssymb]{revtex4}

\begin{document}
\title{Statistical thermodynamics of supercapacitors and blue engines}
\author{Ren\'{e} van Roij}
\affiliation{Institute for Theoretical Physics, Utrecht University, Leuvenlaan 4, 3584 CE Utrecht, The Netherlands}





\maketitle

\section{Introduction}
The scarcity of fresh water, the depletion of fossil fuels, and the ever-increasing demand for electric power are important issues that receive increasing attention in a variety of branches of science and technology. In all three cases nanoporous carbon electrodes, immersed in a liquid medium with charge carriers, are being considered as device elements.  For instance, in carbide-derived carbon electrodes with nanometer-sized pores filled with an ionic liquid,  electric energy can be stored/released through the adsorption/desorption of ionic charges on/from the surface of the charging/discharging electrodes  \cite{chmiola,millersimon,merlet}. Carbon electrodes are also being explored in capacitive devices to harvest sustainable energy from mixing fresh river water with salty sea water \cite{brogioli,brogioli2,rica,sales}. This salinity-gradient energy,  or ``blue" energy, is obtained from charging up a pair of electrodes immersed in sea water (whereby ions adsorb onto the electrodes at a low potential) and discharging them again immersed in fresh water (whereby ions desorb from the electrodes at a higher potential). This capacitive mixing process, with brackish water as a waste product, intercepts the spontaneous diffusion of ions from high to low salinity in much the same way as heat engines intercept the heat flow from hot to cold heat baths;  for typical salt concentrations in river- and sea water these ``blue engines" can produce of the order of 2kJ of energy per liter of river water, in principle even completely reversibly \cite{boon}. The reverse process, which can be seen as a ``blue fridge", is a desalination process in which two volumes of initially brackish water are converted into a volume of fresh water and a volume of brine by charging up the electrodes in one of the volumes (which then desalinates  due to ion adsorption onto the electrodes, at a high potential) and discharging them in the other volume (which then becomes more salty due to the release of the ions from the electrodes, at a low potential) \cite{biesheuvel}. Of course the ``blue fridge" requires a net energy input, and ongoing research questions involve the efficiency and speed of such processes.

In this contribution we will perform a thermodynamic and statistical-mechanical analysis of supercapacitors and blue engines. We will identify direct similarities between the electric work performed by/onto these devices with mechanical work performed by heat engines or consumed by fridges. Moreover, we will identify a number of Maxwell relations. A distinction emerges between the differential capacity at constant ion number and at constant ion chemical potential, directly equivalent to the heat capacity at constant volume and constant temperature. Finally, we will discuss the charge distribution on a (porous) electrode at a given potential. Throughout we make connection with recently published results, although we start off with a quick reminder of ordinary thermodynamics to clarify the analogies.

\section{Thermodynamics of heat engines: a reminder}
We consider a system of internal energy $U$, volume $V$, and entropy $S$. If the number of particles in the system and all other (geometric, dielectric, magnetic, etc.) characteristics are considered fixed, we can write $U=U(S,V)$ such that
\begin{equation}\label{U}
dU=TdS -pdV,
\end{equation}
with temperature $T$ and pressure $p$ of the system defined by
\begin{equation}
T=\left(\frac{\partial U}{\partial S}\right)_V \mbox{ and } p=-\left(\frac{\partial U}{\partial V}\right)_S.
\end{equation}
Eq.(\ref{U}) is a combined formulation of the First and Second Law of Thermodynamics, where $TdS$ is the amount of heat that the system takes up reversibly from a heat bath (also at temperature $T$) and $pdV$ the reversible mechanical work done {\em by} the system (on the environment also at pressure $p$).

Consider the system to reversibly go through a cycle in the $p-V$ plane, such that the final state is identical to the initial state. From the fact that $U$ is a state function, we conclude that $\oint dU=0$, such that the total
mechanical work performed by the system during the cycle can be written as $W_m\equiv\oint pdV=\oint TdS$. Geometrically this means that the work equals the enclosed area in the $p-V$ plane, but also the enclose area of the cycle in the $T-S$ plane. In other words, reversible work of a cyclic heat engine {\em must} be accompanied by heat exchange. For the system to perform a positive amount of work, i.e. for the system to act as a heat engine, it should typically expand at high pressures (and hence at at high temperatures),  thereby taking up heat from baths during (a part of) the expansion, and compress at lower pressures (and hence at lower temperatures), thereby releasing heat into the colder baths during (a part of) the compression.

Two famous examples of heat engines are the Stirling engine and the Carnot engine, for which the working substance is a classical ideal gas (for which $U\propto T\propto pV$) that cycles in a four-fold fashion. In the (idealised) Stirling engine the high-temperature expansion and the low-temperature compression are performed isothermally (such that $dU=0$ and hence $TdS=pdV$), while the cooling and heating parts take place isochorically ($dV=0$ such that $dU=TdS$). In the Carnot cycle,  the expansion and the compression consist both of an isothermal part ($dU=0$) and an adiabatic part ($dS=0$ such that $dU=-pdV$ thereby cooling and heating the gas upon expanding and compressing the gas, respectively.). It is well known that the Carnot engine yields the most efficient conversion of heat into work for given hot and cold heat reservoirs at high and low temperatures $T_h$ and $T_l$, respectively,  with the mechanical work of the Carnot cycle given by $W_m=\Delta T\Delta S$ where $\Delta T=T_h-T_l$ and $\Delta S$ is the entropy extracted from the hot bath during the isothermal expansion and delivered to the cold bath during the isothermal compression. In the $T-S$ plane this Carnot cycle is represented by a rectangular shape at two fixed temperatures and two fixed entropies.

\section{Thermodynamics of electrode-electrolyte systems}
The system of our actual interest here is a macroscopic electrode with total charge $Q$ in contact with a 1:1 electrolyte that contains $N+Q/e$ counterions and $N$ monovalent coions, such that the combined electrode-electrolyte system is charge neutral. Here $e$ is the proton charge. The temperature $T$ is fixed, and the geometric properties of the electrode (e.g. its surface area, its porosity, the volume and the curvature of its pores, etc.) are assumed to be fixed as well. Moreover, in the case of an aqueous electrolyte the water is treated as a structureless dielectric continuum. The Helmholtz free energy of this system can then be written as $F(N,Q)$, where we drop the dependence on the fixed temperature $T$ and the fixed geometric variables for notational convenience. Regardless the functional form of $F$, which depends on the microscopic details of the ion-electrode and ion-ion interactions, we can write the differential of the state function $F$ generally as
\begin{equation}\label{F}
dF=\mu dN + \Psi dQ,
\end{equation}
where we define the ionic chemical potential $\mu$ and the electrostatic potential of the electrode $\Psi$ as
\begin{equation}
\mu=\left(\frac{\partial F}{\partial N}\right)_Q \mbox{ and }\Psi=\left(\frac{\partial F}{\partial Q}\right)_N.
\end{equation}
We note that $\mu$ and $\Psi$, which are the intensive conjugate variables of the extensive variables $N$ and $Q$, respectively, can be seen as equations of state, i.e. $\mu=\mu(N,Q)$ and $\Psi=\Psi(N,Q)$, which we leave unspecified for now as we focus on general and universal thermodynamic properties of the electrode-electrolyte system of interest.

We first note that $\Psi dQ$ in Eq.(\ref{F}) represents the (isothermal and reversible) electrostatic work done {\em on} the electrode-electrolyte system by its environment (which is also at electric potential $\Psi$) when it provides the system with an additional charge $dQ$.   Likewise $\mu dN$ is the (reversible and isothermal) chemical work done {\em on} the system when its environment (at chemical potential $\mu$) provides $dN$ pairs of salt ions. Consider now a reversible and isothermal
{\em cyclic} process due to charging and discharging processes, possibly at various $N$, under the constraint that the initial and the final state of the electrode-electrolyte system are the same. The total electric work performed {\em by} the system during the cycle equals
$W=-\oint \Psi dQ$, i.e. the work performed  is the (negative of the) enclosed area in the $\Psi-Q$ plane.
Typically work is done ($W>0$) by this cyclic "blue engine"  if the electrode is charged at low voltage (which usually implies a high concentration of ions to screen the electrode charge)  and discharged at high voltage (low salt concentrations). By virtue of $F$ being a state function, such that  $\oint dF=0$ and hence  $W=\oint \mu dN$, we indeed find that reversible cyclic blue engines {\em must} be accompanied by ion exchange processes, whereby $W>0$ implies by thermodynamic necessity that ions are to be taken up ($dN>0$) by the system at a high chemical potential (during (a part of) the charging process) and released again ($dN<0$) at a low chemical potential (during (a part of) the discharging).

If one compares Eqs.(\ref{U}) and (\ref{F}) a striking resemblance appears, not only regarding the (free) energy contribution due to the exchange of heat and ions, $TdS$ and $\mu dN$, but also regarding the mechanical and electric work contributions, $-pdV$ and $\Psi dQ$, respectively. In fact, we can make the following mapping:
$U\leftrightarrow F$, $T\leftrightarrow\mu$, $S\leftrightarrow N$, $-p\leftrightarrow\Psi$, and $V\leftrightarrow Q$. This identification of variables respects the symmetry of extensivity and intensivity, connects (de)compressions of a gas with (dis)charging of the electrolyte-immersed electrode, and  implies that isothermal (constant-$T$) and adiabatic (constant-$S$) volume changes in heat engines are analogous to grand-canonical (constant-$\mu$) and canonical (constant-$N$) charging processes of electrolyte-immersed electrodes, respectively.

With this mapping of variables between heat engines and blue engines in mind, it is interesting to note that the blue engine recently developed by Brogioli \cite{brogioli} to harvest salinity gradient energy is actually equivalent to the (idealised) Stirling engine: Brogioli's electrode charging/discharging processes take place at constant $\mu$, just like Stirling's volume changes take place at constant $T$, and Brogioli's flushing processes to exchange river- and sea water (to change the ion chemical potential) at constant electrode charge are equivalent to Stirling's heat exchanges between hot and cold baths (to change the temperature) at constant volume. In addition, this mapping of variables was exploited in Ref.\cite{boon} to construct a conceptual Carnot-like blue-engine. The key difference with the Brogioli cycle is the replacement of the two flushing steps by a constant-$N$ charging and discharging process, whereby the initially salty water desalinates upon electrode charging due to ion adsorption, and the initially fresh water salinates upon discharging due to ion desorption. This Carnot-like cycle composed of two iso-$\mu$ and two iso-$N$ (dis)charging steps, yields an electric work output $W=\Delta N\Delta\mu$ per cycle, where $\Delta N$ is the number of ion pairs that flows during the iso$-\mu$ parts of the cycle from the salty to the fresh water through an adsorption-desorption process onto the electrodes, and where $\Delta\mu$ is the chemical potential difference between the salty and the fresh water. The rectangular shape of the enclosed area of the cycle in the $\mu$-$N$ representation is the hallmark for the most efficient process, since it makes explicit that each of the $\Delta N$ ion pairs contributes its full chemical potential difference $\Delta\mu$ to the total work; more work (per cycle per transferred ion pair) is thermodynamically impossible.

There is, however, one key difference between the nature of the mapped variables $T$ and $\mu$, since $T$ does have a well-defined absolute zero whereas $\mu$ is only defined up to an arbitrary reference potential. For that reason there is
no well-defined analogue of the Carnot heat-engine efficiency $\Delta T/T_h$; the analogous expression $\Delta\mu/\mu_h$ with $\mu_h$ the high chemical potential of the sea water is meaningless.

\section{Maxwell relations and response functions}
Starting from Eq.(\ref{F}) a number of of thermodynamic relations for (super)capacitors immersed in an ionic fluid can be constructed in full analogy to the standard relations that follow from Eq.(\ref{U}) for heat exchange and volume work. In the latter case it proves convenient, for instance, to consider Legendre transformations of $U(S,V)$ to obtain thermodynamic potentials such as the Helmholtz free energy, the Gibbs free energy, or the grand potential, from which a set of Maxwell relations follows by considering second derivatives with respect to two different variables \cite{callen}. What emerges from this is a set of five measurable and therefore interesting response functions associated with heat transfer and pressure-volume work: the constant-volume and constant pressure heat capacities $c_V=T(\partial S/\partial T)_V$ and $c_p=T(\partial S/\partial T)_p$, the isothermal and the adiabatic compressibilities $\kappa_T=-V^{-1}(\partial V/\partial p)_T$ and $\kappa_S=-V^{-1}(\partial V/\partial p)_S$, and the isobaric thermal expansivity $\beta_p=V^{-1}(\partial V/\partial T)_p$. However, standard textbook thermodynamics dictates (on the basis of reciprocity relations and Maxwell relations) that these five quantities are not all independent, as they satisfy
\begin{equation}\label{relation1}
 \frac{c_p}{c_V}=\frac{\kappa_T}{\kappa_S} \hspace{1cm}\mbox{and}\hspace{1cm} c_p-c_V=T\frac{\beta_p^2}{\kappa_T},
\end{equation}
such that there are in fact only {\em three} independent response functions. As a consequence of the second relation together with $\kappa_T>0$,  it is guaranteed that $c_p>c_V$ and hence from the first one that $\kappa_T>\kappa_S$.  Below we follow exactly the same thermodynamic arguments for the electrode-electrolyte system of interest here, building on the mapping of the variables as discussed in the previous section.

We consider three Legendre transformations of $F(N,Q)$, which we denote by the Gibbs-like free energy $G(N,\Psi)=F-\Psi Q$, the grand-potential $\Omega(\mu, Q)=F-\mu N$, and the thermodynamic potential $Y(\mu, \Psi)=F-\mu N-\Psi Q$. Note that all these potentials depend implicitly also on $T$ and on the geometric properties of the electrode (e.g. the volume), which renders even $Y$ well-defined as any bonafide thermodynamic potential must depend on at least one extensive variable. The differentials of these potentials are given by
\begin{eqnarray}
dG&=&\mu dN -Q d\Psi, \label{G}\\
d\Omega&=&-Nd\mu +\Psi dQ,\nonumber\\
dY&=&-Nd\mu-Qd\Psi\nonumber.
\end{eqnarray}
By taking ``off-diagonal" second derivatives of each of the four potentials, the following four Maxwell equations can straightforwardly be derived:
\begin{eqnarray}
\frac{\partial^2 F}{\partial N\partial Q}&=&\left(\frac{\partial\Psi}{\partial N}\right)_Q=\left(\frac{\partial\mu}{\partial Q}\right)_N;\nonumber\\
\frac{\partial^2 G}{\partial N\partial \Psi}&=&-\left(\frac{\partial Q}{\partial N}\right)_\Psi=\left(\frac{\partial\mu}{\partial \Psi}\right)_N;\nonumber\\
\frac{\partial^2 \Omega}{\partial \mu\partial Q}&=&\left(\frac{\partial\Psi}{\partial \mu}\right)_Q=-\left(\frac{\partial N}{\partial Q}\right)_\mu;\label{maxOmega}\\
-\frac{\partial^2 Y}{\partial \mu\partial \Psi}&=&\left(\frac{\partial N}{\partial \Psi}\right)_\mu=\left(\frac{\partial Q}{\partial \mu}\right)_\Psi\equiv\alpha_\Psi.\label{maxY}
\end{eqnarray}
We note that an alternative derivation of Eq.(\ref{maxOmega}) was recently reported, and in fact both sides of the equation as obtained from measurements were successfully compared \cite{rica}, where the explicit ideal-solution relation between chemical potential and salt concentration was used. We also note that $\alpha_\Psi$ as defined in Eq.(\ref{maxY}) plays the same role here as the isobaric expansivity $\beta_p$ in ``standard" thermodynamics.

By considering ``diagonal" second derivatives of the two potentials that depend on $Q$ we define the iso-$\mu$ and the iso-$N$ capacitances $C_N$ and $C_\mu$,
\begin{eqnarray}
\left(\frac{\partial^2 F}{\partial Q^2}\right)^{-1}&=&\left(\frac{\partial Q}{\partial \psi}\right)_N\equiv C_N,\label{CN}\\
\left(\frac{\partial^2 \Omega}{\partial Q^2}\right)^{-1}&=&\left(\frac{\partial Q}{\partial \psi}\right)_\mu\equiv C_{\mu}\label{Cmu},
\end{eqnarray}
which we recognise on the basis of our mapping of variables as the analogues of the two compressibilities.  Taking  ``diagonal"  second derivatives of the the two potentials that depend on $N$ we find
\begin{eqnarray}
\left(\frac{\partial^2 F}{\partial N^2}\right)^{-1}&=&\left(\frac{\partial N}{\partial \mu}\right)_Q\equiv \chi_Q\nonumber\\
\left(\frac{\partial^2 G}{\partial N^2}\right)^{-1}&=&\left(\frac{\partial N}{\partial \mu}\right)_\Psi\equiv \chi_{\Psi},
\end{eqnarray}
where the $\chi$'s play the role of the heat capacities.

The five quantities $C_{\mu}$ , $C_N$, $\chi_{\Psi}$ , $\chi_Q$, and $\alpha_\Psi$ are not all independent, and satisfy relations akin to the two standard relations between $c_p$, $c_V$, $\kappa_S$, $\kappa_T$, and $\beta_p$ of Eq.(\ref{relation1}). Standard thermodynamic manipulations involving the reciprocal and the reciprocity relations yield
\begin{equation}\label{relation2}
\frac{C_{\mu}}{C_N}=\frac{\chi_{\Psi}}{\chi_Q} \hspace{1cm}\mbox{and}\hspace{1cm} \chi_\Psi-\chi_Q=\frac{\alpha^2_\Psi}{C_\mu}.
\end{equation}
With $C_{\mu}>0$, which is a stability requirement as we will see below, we thus find that $\chi_\Psi>\chi_Q$ and $C_{\mu}>C_N$. Note that thermodynamics does {\em not} provide numerical values for these quantities, as this would require a microscopic or molecular theory for the electrode-electrolyte system. The strength of these thermodynamic relations lies, however, in their generality: it is thermodynamically {\em guaranteed} for {\em any} electrode-electrolyte system that $C_\mu>C_N$. It is therefore guaranteed that $\Psi(Q)$ rises faster with $Q$ at fixed $N$ than at fixed $\mu$, the difference being larger if $\alpha_\Psi$ is larger, i.e. if the electrode exhibits a larger adsorption-to-potential or charge-to-concentration response.
Given that a large difference $C_{\mu}-C_N$ gives rise to a large area in the enclosed $\Psi-Q$ plane, and hence a large amount of work during a (Carnot-like) cycle, it could be beneficial for these devices to be based on electrode-electrolyte combinations with a large $\alpha_\Psi$.

\section{Ensembles and charge distribution}
We have been concerned with an electrode-electrolyte system of which the thermodynamics can be described by the Helmholtz free energy $F(N,Q)$. This system has an underlying microscopic Hamiltonian, denoted by ${\cal H}$ here, that depends on all the degrees of freedom of all the ions and all the surface charges. At temperature $T$, the statistical probability of a microscopic configuration is then given by the Boltzmann weight $\exp(-\beta {\cal H})/Z(N,Q)$, where the normalization factor $Z(N,Q)$ is the canonical partition function and where $\beta^{-1}=k_BT$. The Helmholtz free energy, with the differential given in Eq.(\ref{F}), follows as $F(N,Q)=-k_BT\ln Z(N,Q)$ as usual.

If we now consider the electrode-electrolyte system at fixed $N$ and $\Psi$, the total electrode charge $Q$ is a fluctuating quantity that takes values according to the thermal probability distribution
\begin{equation}\label{PQ}
P(Q)=\frac{\exp(-\beta F(N,Q)+\beta\Psi Q)}{{\cal Z}(N,\Psi)},
\end{equation}
where the normalisation factor is  the Gibbs-like partition function ${\cal Z}(N,\Psi)=\sum_{Q}\exp(-\beta F(N,Q)+\beta\Psi Q)\equiv\exp(-\beta G(N,\Psi))$, with $G(N,\Psi)$ the Gibbs-like potential with a differential given by Eq.(\ref{G}).  It readily follows that the average electrode charge is given by $\langle Q\rangle_N =\sum_Q P(Q)Q=k_BT{\cal Z}^{-1}(\partial {\cal Z}/\partial \Psi)_N=-(\partial G/\partial\Psi)_{N}$, in agreement with the differential of Eq.(\ref{G}). Likewise one can write $\langle Q^2\rangle_N=(k_BT)^2{\cal Z}^{-1}(\partial^2 {\cal Z}/\partial \Psi^2)_N$, which after some elementary algebra gives rise to the following identity for the variance
\begin{equation}\label{Qvar}
\langle Q^2\rangle_N-\langle Q\rangle_N^2=k_BTC_N,
\end{equation}
with $C_N$ the constant-$N$ differential capacity defined in Eq.(\ref{CN}). One can also show that the variance of $Q$ at fixed $\mu$ and $\Psi$ is given by $k_BTC_{\mu}$, which is larger than in the iso-$N$ case since $C_{\mu}>C_N$ as we have seen above. Expression (\ref{Qvar}) shows that the differential capacitance can be measured from the fluctuations of the electrode charge, completely equivalently to measurements of the heat capacity from energy fluctuations and the compressibility from volume fluctuations in ``ordinary" $(NVU)$ and $(NVT)$ ensembles, respectively.

Of course the differential capacity $C$ (either $C_N$ or $C_{\mu}$) as well as the average electrode charge $\langle Q\rangle$ (either $\langle Q\rangle_N$ or $\langle Q\rangle_{\mu})$  are extensive quantities that scale linearly with the system size (in this case the electrode area). Therefore the standard deviation of the charge, $(k_BTC)^{1/2}\equiv \delta Q$, becomes much smaller than the average charge $\langle Q\rangle$ for thermodynamically large electrodes. However, our  analysis can also be applied to a small (sub-)system {\em provided} it is large enough to be statistically independent (in fact $F(N,Q)=F(NM,QM)/M$ for $M>1$ must hold). An example includes typical computer simulations of $10^2$-$10^4$ ions near electrode areas of the order of tens of nm$^2$. For such small electrode areas the charge fluctuations at fixed $\Psi$ can be significant, and in fact even be of the order of $\langle Q\rangle$. Consider, for instance, a typical carbon-based supercapacitor with an areal capacity of the order of several $\mu$F/cm$^2$ at a potential $\Psi=1$V, such that the average charge-density is of the order $10^{-2}e$/nm$^{2}$. A patch of electrode of the order of 100 nm$^2$ contains, therefore, a charge $\langle Q\rangle \pm \delta Q$ of the order of  $e\pm e$, indicating that a significant fraction of nm-sized patches carries a charge that is opposite to the average charge.

In fact it is possible to calculate the complete charge distribution $P(Q)$ as defined in Eq.(\ref{PQ}) from the formalism that we used here by expanding $F(N,Q)$ about  the most-probable charge $Q^*$, defined by $P'(Q^*)=0$, i.e. by $F'(N,Q^*)=\Psi$, where a prime denotes a derivative with respect to $Q$. Skipping the $N$-depence for notational convenience, we then find
\begin{eqnarray}
P(Q)&\propto& \exp\left(-\frac{1}{2}(Q-Q^*)^2 F''(Q^*)\right. \nonumber\\
&&\,\,\,\,\,\,\,\,\,\,\,\,\,\,\,\,+\left. \frac{1}{6}(Q-Q^*)^3F'''(Q^*) +\dots\right)\nonumber\\
&=&\exp\left(-\frac{(Q-Q^*)^2}{2k_BT C^*} + \frac{(Q-Q^*)^3 C'(Q^*)}{6 k_BT (C^*)^2} +\dots\right),\nonumber\\\label{PQ2}
\end{eqnarray}
where $C^*=C_N(Q^*)$. On the basis of extensivity arguments the cubic and higher-order terms in the exponent can be ignored in the thermodynamic limit, such that $P(Q)$ is a Gaussian with $Q^*=\langle Q\rangle$ and a variance in accordance with Eq.(\ref{Qvar}). However, for smaller systems the higher order terms may be relevant, at least in the case when $C'(Q)=(\partial C/\partial Q)_N\neq 0$, i.e. when the differential capacity depends significantly on the average charge (and hence on the applied voltage) \cite{hatlo}. Eq.(\ref{PQ2}) appears to be in agreement with the findings in Ref.\cite{merlet} on graphite electrodes, which give an almost Gaussian charge distribution and an essentially vanishing $C'$. However, the significantly skewed charge distributions as found Ref.\cite{merlet} for nanoporous carbide-derived carbon electrodes require some further attention, as the capacity is reported to be essentially constant, which should yield a Gaussian distribution according to the present derivation. We speculate that the individual carbon atoms of the electrode may be too small to be viewed as a statistically independent subsystem, although more research is needed to clarify this point.

\section{Summary}
We compare the differential of the Helmholtz free energy $F(N,Q)$ of an electrode-electrolyte system with that of the energy $U(S,V)$ of an ``ordinary" thermal system, and identified the transferred heat $TdS$ with the  ion flow contribution $\mu dN$, and the mechanical work $-pdV$ with the electric work $\Psi dQ$. By a mapping of the variables $(N,\mu,Q,\Psi)$ of present interest onto $(S,T,V,-p)$, we can define Legendre transformations of $F(N,Q)$, identify Maxwell relations, and formulate Eq.(\ref{relation2}) analogous to (\ref{relation1}). With this mapping we identify the Brogioli blue engine as a Stirling heat engine, and we discuss a Carnot-like blue engine that should have the optimal conversion of mixing entropy to work. Finally we discuss the charge distribution at fixed electrode potential, and show that the variance of the charge scales with the differential capacitance.

\section{Acknowledgement}
It is a pleasure to thank Niels Boon, Benjamin Rotenberg, Doriano Brogioli, Raul Rica, Francesco Mantegazza, and Maarten Biesheuvel for inspiration and useful discussions.


\begin{thebibliography}{99}
\bibitem[{Chmiola {\em et al.} (2006)}]{chmiola} J. Chmiola, G. Yushun, Y. Gogotsi, C. Portet, P. Simon, and P.L. Taberna,
Science {\bf 313}, 1760 (2006).
\bibitem[{Miller and Simon (2008)}]{millersimon} J.R. Millar and P. Simon, Science {\bf 321}, 651 (2008).
\bibitem[{Merlet {\em et al.} (2012)}]{merlet}
C. Merlet, B. Rotenberg, P.A, Madden, P.-L. Taberna, P. Simon, Y. Gogotsi, and M. Salanne, Nature Materials {\bf 11}, 306 (2012).
\bibitem[{Brogioli (2009)}]{brogioli} D. Brogioli,  Phys. Rev. Lett. {\bf 103}, 058501 (2009).
\bibitem[{Brogioli {\em et al.} (2011)}]{brogioli2} D. Brogioli, R. Zhao,; P.M. Biesheuvel, Energy Environm. Sci. {\bf 4}, 772 (2011).
\bibitem[{Rica {\em et al.} (2012)}]{rica} R.A. Rica, R. Ziano, D. Salerno, F. Mantegazza, and D. Brogioli, Phys. Rev. Lett. 156103 {\bf 109} (2012).
\bibitem[{Sales {\em et al.} (2010)}]{sales} B.B. Sales, M. Saakes, J. Post, C.J.N. Buisman, P.M. Biesheuvel, and H.V.M. Hamelers, Env. Sci. Techn. {\bf 44}, 5661 (2010).
\bibitem[{Boon and van Roij (2011)}]{boon} N. Boon and R. van Roij, Mol. Phys. {\bf 109}, 1229 (2011).
\bibitem[{Biesheuvel (2009)}]{biesheuvel} P.M. Biesheuvel, J. Colloid Interface Sci. {\bf 332}, 258 (2009); S. Porada et al., Appl. Mater. Interfaces {\bf 4}, 1194 (2012).
\bibitem[{Callen (1985)}]{callen} H.B. Callen, {\em Thermodynamics and an introduction to thermostatistics},  John Wiley and Sons (1985).
\bibitem[{Hatlo {\em et al.} (2012)}]{hatlo} M.M. Hatlo, R. van Roij and L. Lue, Europhys. Lett. {\bf 97}, 28010 (2012).
\end{thebibliography}
\end{document}